\documentclass[12pt]{article}
\begin{document}
\vspace*{-.6in}
\thispagestyle{empty}
\begin{flushright}
CALT-68-2402
\end{flushright}
\baselineskip = 18pt

\vspace{1.5in} {\Large
\begin{center}
COMMENTS ON SUPERSTRING INTERACTIONS IN A PLANE-WAVE
BACKGROUND\end{center}} \vspace{.5in}

\begin{center}
John H. Schwarz
\\
\emph{California Institute of Technology\\ Pasadena, CA  91125, USA}
\end{center}
\vspace{1in}

\begin{center}
\textbf{Abstract}
\end{center}
\begin{quotation}
\noindent The three string vertex for Type IIB superstrings in a
maximally supersymmetric plane-wave background is investigated.
Specifically, we derive a factorization theorem for the Neumann
coefficients that generalizes a flat-space result that was
obtained some 20 years ago. The resulting formula is used to
explore the leading large $\mu$ asymptotic behavior, which is
relevant for comparison with dual gauge theory results.
\end{quotation}

\newpage

\pagenumbering{arabic}

\section{Introduction}

It was recently discovered that Type IIB superstring theory admits
a maximally supersymmetric plane-wave background
\cite{Blau:2001ne}. Moreover, the string theory in this background
is tractable, despite the fact that the background contains a
nonvanishing RR field, in the light-cone GS formalism
\cite{Metsaev:2001bj}. In that approach the world-sheet theory
consists of free massive bosons and fermions. Subsequently, there
was a proposal to relate the string states and their interactions
to holographically dual calculations in terms of certain limiting
operators and their correlation functions in ${\cal N} = 4$ super
Yang-Mills theory \cite{Berenstein:2002jq}.

The string interactions are encoded (using the language of
light-cone-gauge string field theory) in a cubic interaction
vertex. This vertex has been formulated in a pair of very nice
papers by Spradlin and Volovich \cite{Spradlin:2002ar}
\cite{Spradlin:2002rv}. Their work generalizes the flat-space
light-cone-gauge field theory results of \cite{Green:1982tc}
\cite{Green:hw} to the plane-wave geometry.

This paper has two goals. First, we wish to make the formulas for
the Neumann coefficients that enter in the interaction vertex more
explicit. Specifically, one wants explicit formulas for the
inverse of a certain infinite-dimensional matrix. We succeed in
expressing the inverse matrix in a factorized form in terms of a
certain infinite component vector, but we have not yet obtained
explicit formulas for the vector. However, even this step is
useful for our second goal: exploring the large $\mu$ (large
curvature) limit of the geometry, which is required for making
contact with perturbative gauge theory computations.

\section{Review of basic formulas}

The type IIB superstring in the maximally supersymmetric
plane-wave background is described in light-cone gauge by a free
world-sheet theory. The eight bosonic and eight fermionic
world-sheet fields each have mass $\mu$, a parameter that enters
in the description of the plane-wave geometry and the RR five-form
field strength. The mass term has two important consequences. One
is that it leads to a mixing of left-movers and right-movers. The
other is that the zero modes are also described by harmonic
oscillators of finite frequency. Altogether, a convenient labeling
of the bosonic lowering and raising operators arising from
quantization of the free world-sheet theory is $a_m^I$ and $a_m^{I
\dagger}$, where $m$ runs from minus infinity to plus infinity and
$I = 1,\ldots,8$. These satisfy ordinary oscillator commutation
relations
\begin{equation}
[a_m^I, a_n^{J \dagger}] = \delta_{mn} \delta^{IJ}.
\end{equation}
There are also fermionic oscillators $b_m^{\alpha}$ and
$b_m^{\alpha \dagger}$, which will not be discussed in this paper.

The spectrum of the free string theory is described by the
light-cone Hamiltonian
\begin{equation}
H_2 = \sum_{m= -\infty}^{\infty} \omega_m N_m
\end{equation}
where $N_m$ is the number of excitations of level $m$ oscillators
\begin{equation}
N_m = \sum_{I=1}^8 a_m^{I \dagger} a_m^I + {\rm fermionic\, terms}
\end{equation}
where the frequencies are given by
\begin{equation}
\omega_m = \sqrt{m^2 + \mu^2 \alpha^2}.
\end{equation}
The second term in the square root is actually $(\alpha' \mu
p_-)^2$, but we set the Regge slope $\alpha'=2$ and define $\alpha
= 2 p_-$. (In flat space, we used the symbol $p^+$ rather than
$p_-$, but in curved space a lower index is preferable.) The
physical spectrum is given by the product of all the oscillator
spaces subject to one constraint
\begin{equation}
\sum_{m= -\infty}^{\infty} m N_m =0.
\end{equation}
In flat space this constraint reduces to the usual level-matching
condition for left-movers and right-movers.

The three-string interaction vertex for type IIB superstrings in
flat space was worked out in \cite{Green:1982tc} and
\cite{Green:hw} and generalized to the plane-wave geometry in
\cite{Spradlin:2002ar} and \cite{Spradlin:2002rv}. The formula can
be written rather elegantly in terms of functionals, but to make
its meaning precise it is desirable to expand it out in terms of
oscillators. A convenient notation is to use a tensor product of
three string Fock spaces, labeled by an index $r=1,2,3$. Then the
three string interaction vertex contains a factor
\begin{equation}
|\, V_B> = {\rm exp} \left( {1\over 2} \sum_{r,s =1}^3 \sum_{m,n
=-\infty}^{\infty} \sum_{I=1}^8 a_{mr}^{I\dagger} \bar N_{mn}^{rs}
a_{ns}^{I\dagger}\right) |\, 0>.
\end{equation}
The quantities $\bar N_{mn}^{rs}$, called Neumann coefficients,
are the main objects of concern in this paper. Their definition
here differs from that used in \cite{Green:1982tc} and
\cite{Green:hw} by factors of $\sqrt{m n}$. The definition given
here is more natural for the $\mu\neq 0$ generalization. The three
string vertex also contains a similar expression $|\, V_F>$ made
out of the fermionic oscillators and a ``prefactor'' that is
polynomial in the various oscillators. We will not consider either
of these in this paper.

In describing the Neumann matrices, it is convenient to consider
separately the cases in which each of the indices $m,n$ are either
positive, negative or zero. Henceforth, the symbols $m,n$ will
always denote positive integers. Negative integers will be
indicated by displaying an explicit minus sign.  One result of
\cite{Spradlin:2002ar}, for example, using matrix notation for the
blocks with positive indices, is
\begin{equation}
\bar N^{rs} = 1 - 2 (C_r C^{-1})^{1/2} A^{(r)T}
\Gamma_+^{-1}A^{(s)} (C_s C^{-1})^{1/2}.
\end{equation}
Here $C_{mn} = m \delta_{mn}$ and $(C_r)_{mn} = \omega_{rm}
\delta_{mn}$, where $ \omega_{rm} = \sqrt{ m^2 +
(\mu\alpha_r)^2}$. These are simple diagonal matrices. The great
challenge is to understand the rest of the formula. The
definitions of $A^{(r)}$ and $\Gamma_+$, and other expressions
that appear here, are collected in the appendix.

The blocks with both indices negative are related in a simple way
to the ones with both indices positive by
\begin{equation}
\bar N_{-m-n}^{rs} = - \left( U_r \bar N^{rs} U_s \right)_{mn},
\end{equation}
where
\begin{equation}
U_r = C^{-1} (C_r - \mu \alpha_r).
\end{equation}
In the case of $\bar N^{33}$ these are the only nonvanishing
terms. For the other Neumann coefficients the other nonvanishing
terms are
\begin{equation}
\bar N_{m0}^{3r} = \bar N_{0m}^{r3} = \sqrt{2\mu\alpha_r}\,
\epsilon^{rs}\alpha_s \left[ \left(C_3 C^{-1}\right)^{1/2}
\Gamma_+^{-1} B \right]_m \quad r,s = 1,2
\end{equation}
\begin{equation}
\bar N_{m0}^{rs} = \bar N_{0m}^{sr} = \sqrt{2\mu\alpha_s}\,
\epsilon^{st}\alpha_t \left[ \left(C_r C^{-1}\right)^{1/2}
A^{(r)T} \Gamma_+^{-1} B \right]_m \quad r,s,t = 1,2
\end{equation}
and
\begin{equation}
\bar N_{00}^{rs} = \delta^{rs} + {\sqrt{\alpha_r\alpha_s} \over
\alpha_3} -\mu\sqrt{\alpha_r\alpha_s}\,
\epsilon^{rt}\epsilon^{su}\alpha_t \alpha_u B^T \Gamma_+^{-1} B
 \quad r,s,t,u = 1,2
\end{equation}
\begin{equation}
\bar N_{00}^{3r} = \bar N_{00}^{r3} = - \sqrt{-{\alpha_r \over
\alpha_3}} \quad r = 1,2
\end{equation}
The asymmetry between string number three and the other two
strings is a reflection of the fact that the $\mu$ dependence of
the formula breaks the cyclic symmetry that is present in the flat
space case. Evidently, crossing symmetry no longer holds in the
plane-wave geometry.

To make the formulas useful for comparison with the dual gauge
theory, it would be helpful to have explicit formulas in which the
various matrix multiplications and inversions have been
analytically evaluated. The quantities that we especially would
like to evaluate explicitly are the matrix $\Gamma_+^{-1}$, the
vector
\begin{equation}
Y =\Gamma_+^{-1} B,
\end{equation}
and the scalar
\begin{equation}
k = B^T \Gamma_+^{-1} B.
\end{equation}
In the case of flat space ($\mu
=0$) the results are known. Specifically
\begin{equation} \label{factorize}
\bar N_{mn}^{rs} = - {mn\alpha \over m\alpha_s + n \alpha_r} \bar
N_m^r \bar N_n^s \quad {\rm for}\,  \mu =0
\end{equation}
where $\alpha = \alpha_1 \alpha_2 \alpha_3$ and
\begin{equation} \label{Nrflat}
\bar N_m^r = {\sqrt{m} \over \alpha_r} f_m(-\alpha_{r+1}/\alpha_r)
e^{m\tau_0 /\alpha_r} \quad {\rm for}\,  \mu =0,
\end{equation}
\begin{equation}
f_m(\gamma) = {\Gamma(m\gamma) \over m! \Gamma (m\gamma +1 -m)}
\end{equation}
and
\begin{equation}
\tau_0 = \sum_{r=1}^3 \alpha_r {\rm log} |\alpha_r|
\end{equation}
In particular, still for $\mu =0$, $\Gamma_+^{-1} = {1\over 2} (1
- \bar N^{33})$, $Y_m = - \bar N_m^3$, and $k = 2\tau_0/\alpha$.

\section{Neumann coefficients for $\mu \neq 0$}

In this section we will derive the generalization of
eq.(\ref{factorize}) that holds for the plane-wave geometry. The
method of derivation is a fairly straightforward generalization of
the one used for flat space in \cite{Green:1982tc}. We begin by
defining
\begin{equation} \label{tildeG}
\tilde\Gamma_+ = \sum_{r=1}^3 A^{(r)} U_r^{-1} A^{(r)T}
\end{equation}
and then considering the product
\begin{equation}
\Gamma_+\,  C^{-1}\,  \tilde\Gamma_+ = (U_3 + \sum_1^2 A^{(r)} U_r
A^{(r)T})\, C^{-1}\, (U_3^{-1} + \sum_1^2 A^{(s)} U_s^{-1}
A^{(s)T})
\end{equation}
Using various identities given in the appendix, this simplifies to
\begin{equation}
\Gamma_+\,  C^{-1}\,  \tilde\Gamma_+ = U_3 C^{-1} \tilde\Gamma_+ +
\Gamma_+ C^{-1} U_3^{-1} -{1\over 2} \alpha_1 \alpha_2 B B^T.
\end{equation}
The next step is to use eqs.~(\ref{Uformula}) and (\ref{AAT2}) to
deduce that
\begin{equation}
\tilde\Gamma_+ = \Gamma_+ + \mu \alpha BB^T
\end{equation}
Substituting this into the previous equation and multiplying left
and right by $\Gamma_+^{-1}$ gives
\begin{equation} \label{idwithZ}
C^{-1} U_3^{-1} \Gamma_+^{-1} + \Gamma_+^{-1} U_3 C^{-1} = C^{-1}
+ {1\over2} \alpha_1 \alpha_2 Y Y^T + \mu \alpha Z Y^T
\end{equation}
where we have defined
\begin{equation}
Z = (1 -\Gamma_+^{-1} U_3) C^{-1} B.
\end{equation}

The next step is to eliminate $Z$ from eq.~(\ref{idwithZ}). This
is achieved by multiplying the equation on the right with the
vector $B$. This gives a linear equation for $Z$, whose solution
is
\begin{equation}
Z = {1\over 1 + \mu \alpha k}(C^{-1} U_3^{-1} - {1\over 2}
\alpha_1 \alpha_2 k)  Y.
\end{equation}
Substituting this back into eq.~(\ref{idwithZ}) and simplifying
gives the formula
\begin{equation} \label{symformula}
\{ \Gamma_+^{-1} , C_3\} = C + {1\over2} {\alpha_1 \alpha_2 \over
1 + \mu \alpha k} C U_3^{-1} Y Y^T CU_3^{-1}.
\end{equation}
If we had explicit formulas for the vector $Y$ and the scalar $k$,
this formula would give us the matrix $\Gamma_+^{-1}$. It can be
recast as a formula for the Neumann coefficient matrix $\bar
N^{33}_{mn}$. The result is
\begin{equation}
\bar N^{33}_{mn} = - {mn\alpha_1 \alpha_2 \over 1 + \mu \alpha k}
{\bar N_m^3 \bar N_n^3 \over \omega_{3m} + \omega_{3n}}
\end{equation}
where
\begin{equation}
\bar N_m^3 = -\left[ (C^{-1}C_3)^{1/2} U_3^{-1} Y\right]_m.
\end{equation}
Some further simple manipulations give the generalization
\begin{equation}
\bar N^{rs}_{mn} = - {mn\alpha \over 1 + \mu \alpha k} {\bar N_m^r
\bar N_n^s \over \alpha_s\omega_{rm} + \alpha_r\omega_{sn}}
\end{equation}
where
\begin{equation}
\bar N_m^r = -\left[ (C^{-1}C_r)^{1/2} U_r^{-1} A^{(r)T}
Y\right]_m.
\end{equation}
This is the desired generalization of the flat-space formula
eq.~(\ref{factorize}). However, we are still lacking a
generalization of the explicit formula (\ref{Nrflat}) as well as
an explicit formula for $k$.

\section{An Involution}

We are primarily interested in the case in which the mass
parameter $\mu$ is positive. In particular, in the next section we
will explore asymptotic properties for $\mu$ large and positive.
The formulas for $\mu$ large and negative are different. We can be
quite explicit about this by relating the two cases. For this
purpose we define $\tilde\Gamma_+ = \Gamma_+(-\mu)$, $\tilde Y=
Y(-\mu)$, and $\tilde k = k(-\mu)$. The expression
$\tilde\Gamma_+$ was already introduced in eq.~(\ref{tildeG}). In
fact, we found that
\begin{equation}
\tilde\Gamma_+ = \Gamma_+ + \mu \alpha BB^T.
\end{equation}
This equation can be inverted to give
\begin{equation}
\tilde\Gamma_+^{-1} = \Gamma_+^{-1} - {\mu \alpha \over 1 + \mu
\alpha k} YY^T.
\end{equation}
Multiplying by $B$ on the right one deduces that
\begin{equation}
\tilde Y = {1 \over 1 + \mu \alpha k} Y
\end{equation}
and
\begin{equation}
\tilde k = {k \over 1 + \mu \alpha k} .
\end{equation}
These formulas are of interest because they allow for nontrivial
checks of various formulas. Each one must transform into another
correct equation under the transformation $\mu \to - \mu$. We have
checked this in every case, and no new equations beyond those
already presented are generated in this way.

\section{Large $\mu$ asymptotics}

In the duality between string theory and gauge theory, it is
necessary to consider large $\mu$ in order to make contact with
perturbative gauge theory calculations. Therefore in this section
we shall work out some of the leading terms in the asymptotic
expansions of $\Gamma_+^{-1}$, $Y = \Gamma_+^{-1}B$, and $k = B^T
Y =B^T\Gamma_+^{-1}B$. Some preliminary studies of these
expansions, which we will extend, have been made previously in
\cite{Huang:2002wf} \cite{Spradlin:2002rv} \cite{Klebanov:2002mp}.

As will become evident, the leading (large $\mu$) term in the
expansion of $\Gamma_+^{-1}$ is given by the first term on the
right-hand side of eq.~(\ref{symformula}). Therefore let us
extract this term by defining
\begin{equation} \label{decomp}
\Gamma_+^{-1} = {1 \over 2} C C_3^{-1} + R
\end{equation}
The first term has the expansion
\begin{equation}
{1 \over 2} C C_3^{-1} \to - {C \over 2 \mu \alpha_3} + {C^3 \over
4 (\mu \alpha_3)^3} + \dots
\end{equation}
We will find that the leading term in $R$ is of order $\mu^{-4}$.

To analyze the asymptotic behavior for large positive $\mu$, let
us begin by inserting eq.~(\ref{decomp}) into
eq.~(\ref{symformula}). This gives
\begin{equation} \label{Reqn}
\{R,C_3\} = {1\over 2} {\alpha_1\alpha_2 \over 1 + \mu \alpha k}
CU_3^{-1} Y Y^T CU_3^{-1}.
\end{equation}
We now making an ansatz for the large $\mu$ structure of $R$, with
an unknown coefficient. Specifically, let us try
\begin{equation}
R \to a_R {\pi \over (\mu \alpha_3)^4} \left( {\alpha_1 \alpha_2
\over \alpha_3}\right)^2 {C^3}B B^T {C^3} + \ldots
\end{equation}
This term is of order $\mu^{-4}$. The next term in the expansion
would be of order $\mu^{-6}$. In similar fashion one can argue
that at large $\mu$
\begin{equation}
k =B^T Y \to - {1 \over \mu \alpha}  - {a_k \over \pi (\mu
\alpha_1 \alpha_2)^2} +\ldots
\end{equation}
Inserting these expansions into eq.~(\ref{Reqn}), one learns that
\begin{equation}
a_R a_k = {1 \over 64}.
\end{equation}
While this relation is easy to derive, it is much more difficult
to determine $a_R$ and $a_k$ separately.\footnote{In the first
version of this paper I claimed to prove that $a_R =a_k =x=1/8$.
However, this is not correct. I am grateful to the authors of
\cite{Klebanov:2002mp} for bringing the error to my attention.}

Let us consider now the asymptotic expansion of $Y$. In view of
the above equations, it should have the structure
\begin{equation}\label{Yasym}
Y = \Gamma_+^{-1} B \to {1 \over \mu \alpha_3} \left[ - {1\over 2}
C + \left( {1 \over 4} - x\right) {C^3 \over (\mu \alpha_3)^2} +
\ldots \right] B.
\end{equation}
Note that the leading term is of order $\mu^{-1}$ and the first
correction is of order $\mu^{-3}$.

The value of $x$ is of particular interest, since a certain dual
field theory calculation gives a result that is proportional to
${1\over 2} - 4x$. Based on the light-cone field theory formulas
described here, it was estimated numerically to be approximately
1/16 in \cite{Klebanov:2002mp}, and that is presumably correct.
The field theory analysis of \cite{Chu:2002pd} gave a nonvanishing
result corresponding to the value $x=0$. Daniel Freedman informs
me that he has repeated the field theory calculation, using a
different regularization method, and that he finds a vanishing
result, corresponding to the value $x=1/8$. I hope to obtain an
exact analytic expression for $Y$ from which one could read off
the exact value of $x$, but as yet this has not been achieved.

\section*{Acknowledgments}
I wish to acknowledge helpful discussions with D. Freedman, I.
Klebanov, M. Spradlin, and A. Volovich. I am also grateful to C.
Callan, W. Lee, T. McLoughlin, I. Swanson, and X. Wu for their
collaboration on related topics. The hospitality of the Newton
Institute for Mathematical Sciences and the Aspen Center for
Physics, where portions of this work were carried out, is also
appreciated. This work was supported in part by the U.S. Dept. of
Energy under Grant No. DE-FG03-92-ER40701.

\section*{Appendix}

The light cone momenta in the three-string vertex are proportional
to $\alpha_r$, where we take $\alpha_1, \alpha_2 > 0$, $\alpha_3 <
0$ and $\alpha_1 + \alpha_2 + \alpha_3 =0$. We also define $\beta
= \alpha_1 / \alpha_3$, which satisfies $ -1 < \beta < 0$.

The matrices $A^{(r)}_{mn}$, which appear in the Neumann
coefficients, are given by
\begin{equation}
A^{(1)}_{mn} = {2 \over \pi} (-1)^{m+n+1}  \sqrt{m n} {\beta \,
{\rm sin} m \pi \beta \over n^2 - m^2 \beta^2},
\end{equation}
\begin{equation}
A^{(2)}_{mn} = {2 \over \pi} (-1)^{m+1}  \sqrt{m n} {(\beta +1)\,
{\rm sin} m \pi \beta \over n^2 - m^2 (\beta +1)^2},
\end{equation}
and $A^{(3)}_{mn} = \delta_{mn}$. The indices $m,n$ range from 1
to infinity. Additional quantities that we need are
\begin{equation}
B_m = {2 \alpha_3 \over \pi \alpha_1 \alpha_2} (-1)^{m+1} { {\rm
sin} m \pi \beta \over m^{3/2} }
\end{equation}
and
\begin{equation}
C_{mn} = m \delta_{mn}.
\end{equation}
Additional matrices that involve the mass parameter $\mu$ of the
plane-wave geometry are
\begin{equation}
(C_r)_{mn} = \omega_{rm} \delta_{mn} = \sqrt{m^2 + \mu^2
\alpha_r^2}\, \delta_{mn}
\end{equation}
and
\begin{equation}
U_r = C^{-1} (C_r -\mu \alpha_r).
\end{equation}
Note that
\begin{equation} \label{Uformula}
U_r^{-1} = C^{-1} (C_r + \mu \alpha_r) = U_r + 2 \mu \alpha_r
C^{-1}.
\end{equation}

A crucial construct is the infinite matrix
\begin{equation}
\Gamma_+ = \sum_{r=1}^3 A^{(r)} U_r A^{(r)T}.
\end{equation}
Explicit formulas for its inverse are a main goal of our work.
Related quantities that also are needed are the infinite vector
\begin{equation}
Y = \Gamma_+^{-1} B
\end{equation}
and the scalar
\begin{equation}
k = B^T \Gamma_+^{-1} B.
\end{equation}

The infinite matrices $A^{(r)}_{mn}$ and the infinite vector $B_m$
satisfy a number of useful relations which we record here
\begin{equation}
A^{(r) T} C A^{(s)} = - {\alpha_3 \over \alpha_r}\, C \delta^{rs}
\quad r,s = 1,2
\end{equation}
\begin{equation}
A^{(r) T} C^{-1} A^{(s)} = - {\alpha_r \over \alpha_3}\, C^{-1}
\delta^{rs} \quad r,s = 1,2
\end{equation}
The symbol $T$ means matrix transpose.
\begin{equation}
B^T C B = {2\over \alpha_1 \alpha_2}
\end{equation}
\begin{equation}
B^T C^{-1} B = {2 \pi^2 \over 3 \alpha_3^2}
\end{equation}
\begin{equation}
B^T {C^3 \over C^2 -\lambda^2} B = {1 \over \pi} \left( {\alpha_3
\over \alpha_1 \alpha_2} \right)^2 {{\rm cos} \pi \lambda
(1+2\beta) - {\rm cos} \pi \lambda \over \lambda\, {\rm sin} \pi
\lambda}
\end{equation}
Substituting the special value $\lambda = i \mu \alpha_3$, the
last identity can be recast as
\begin{equation}
B^T {C^3 \over C_3^2} B = -{2 \over \pi\mu\alpha_3} \left(
{\alpha_3 \over \alpha_1 \alpha_2} \right)^2\left( {\rm coth} \pi
\mu \alpha_1 + {\rm coth} \pi \mu \alpha_2 \right)^{-1}.
\end{equation}
Note that the large $\mu$ asymptotic behavior is very sensitive to
the direction in which infinity is approached. In particular, if
it is approached in the positive direction, which is the usual
case of interest, the last expression reduces to $-{1 \over
\pi\mu\alpha_3} \left( {\alpha_3 \over \alpha_1 \alpha_2}
\right)^2$ with exponential precision.

Some additional useful identities are
\begin{equation}
\sum_{r=1}^3 {1 \over \alpha_r} A^{(r)} C A^{(r)T} =0,
\end{equation}
\begin{equation}\label{AAT2}
\sum_{r=1}^3 \alpha_r A^{(r)} C^{-1} A^{(r)T} = {\alpha \over 2} B
B^T,
\end{equation}
where we have introduced
\begin{equation}
\alpha = \alpha_1 \alpha_2 \alpha_3.
\end{equation}

\end{document}